\documentclass[letterpaper]{article} 
\usepackage{aaai2027}  
\usepackage[hyphens]{url}  
\usepackage{graphicx} 
\usepackage{natbib}  
\usepackage{caption} 
\usepackage{algorithm}
\usepackage{algorithmic}
\usepackage{multirow}

\usepackage[table]{xcolor}
\usepackage{adjustbox}
\usepackage{array}
\usepackage{tabularx}
\newcolumntype{L}[1]{>{\raggedright\arraybackslash}p{#1}}
\newcolumntype{C}[1]{>{\centering\arraybackslash}p{#1}}
\newcolumntype{R}[1]{>{\raggedleft\arraybackslash}p{#1}}
\newcolumntype{Y}{>{\centering\arraybackslash}X}
\usepackage{newfloat}
\usepackage{listings}
\DeclareCaptionStyle{ruled}{labelfont=normalfont,labelsep=colon,strut=off} 
\floatstyle{ruled}
\newfloat{listing}{tb}{lst}{}
\floatname{listing}{Listing}

\usepackage{booktabs}
\usepackage{makecell}
\usepackage{colortbl}
\usepackage{pifont}
\usepackage{amsmath}
\usepackage{amssymb}
\nocopyright

\title{CAD: Conflict-Aware Decoding to Mitigate Cross-Modal Hallucinations in
Omnimodal Large Language Models}

\author{
Yuchen Deng\textsuperscript{\rm 1,2},
Chang Sun\textsuperscript{\rm 3},
Hai-Tao Zheng\textsuperscript{\rm 1,2},
Feidiao Yang\textsuperscript{\rm 2},
Yuxing Han\textsuperscript{\rm 1}\corresponding
}

\affiliations{
\textsuperscript{\rm 1}Tsinghua Shenzhen International Graduate School,
Tsinghua University\\
\textsuperscript{\rm 2}Pengcheng Laboratory;\quad
\textsuperscript{\rm 3}School of Cyber Science and Engineering,
Zhengzhou University\\[2pt]
{\color{blue}\url{https://github.com/conviction6/CAD}}
}

\begin{document}

\maketitle

\begin{abstract}
Omnimodal large language models (Omni-LLMs) integrate audio, video, and text, yet remain vulnerable to cross-modal hallucinations, where one modality improperly influences predictions about another. Existing training-free decoders modulate modality influence through perturbation or relevance weighting, but do not assess predictive compatibility within the joint audio-visual branch. Because joint-branch discrepancies may indicate either harmful interference or useful complementarity, reliable intervention requires assessing both discrepancy magnitude and actionability. To this end, we propose Conflict-Aware Decoding (CAD), a training-free framework comprising Potential Conflict Magnitude Estimation (PCME) and Conflict Actionability Assessment (CAA). PCME quantifies potential conflict using audio-video disagreement and the deviation of the joint prediction from a relevance-weighted unimodal reference. CAA then applies Dempster-Shafer reliability discounting to task-space answer relations, using query relevance and answer decisiveness to determine whether intervention is warranted. When an actionable conflict is identified, CAD selectively reallocates decoding weight from the joint branch to the unimodal branches. Experiments on CMM, AVHBench, WorldSense, and VideoMME show that CAD consistently outperforms the base decoder and competitive training-free methods across multiple audio-visual backbones. On Qwen2.5-Omni-7B, CAD improves overall accuracy by \(14.1\) and \(8.0\) percentage points on CMM and AVHBench, respectively, without model retraining.

\end{abstract}

\section{Introduction}
\label{sec:introduction}

\begin{figure}[t]
    \centering
    \includegraphics[width=1.0\linewidth]{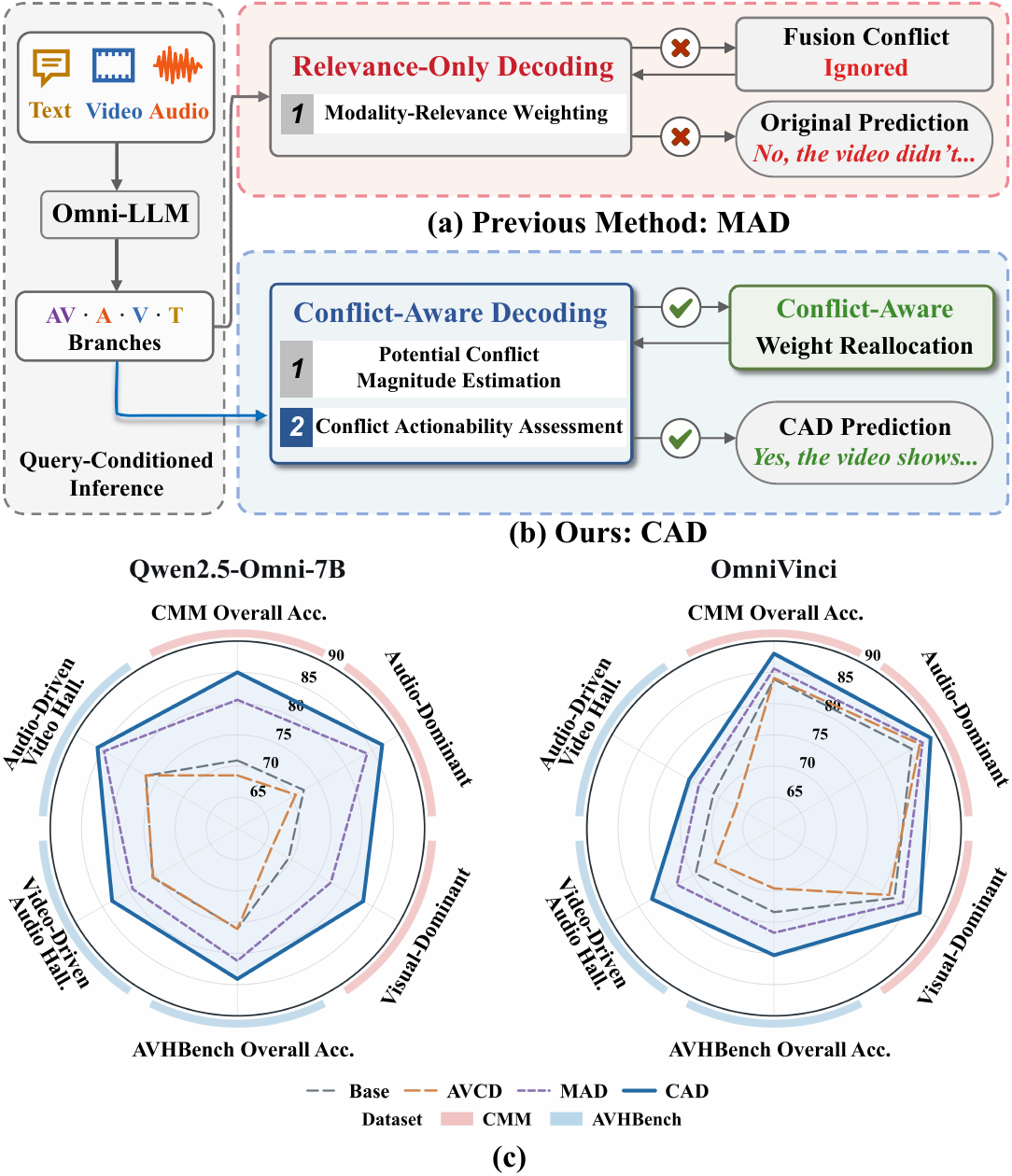}
    \vspace{-4mm}
    \caption{(a) MAD relies on modality relevance alone and overlooks cross-modal conflicts. (b) CAD assesses conflict magnitude and actionability, then reallocates decoding weights to suppress unreliable fusion. (c) On CMM and AVHBench, CAD consistently outperforms Base, AVCD, and MAD with both Qwen2.5-Omni-7B and OmniVinci.}
    \label{fig:Introduction}
    \vspace{-4mm}
\end{figure}

Omnimodal large language models (Omni-LLMs) jointly model text, vision, and audio within a unified framework, enabling integrated multimodal perception and reasoning. Omni-LLMs are now widely applied in complex tasks~\cite{geng2025longvale,liu2024valor} such as audio-visual scene understanding and interactive assistance.

Despite significant progress, Omni-LLMs still face a critical challenge: cross-modal hallucination, which substantially undermines their reliability and stability in practical applications. Effective cross-modal fusion can exploit complementary audio and visual evidence, whereas poorly coordinated fusion may introduce harmful interference. Unlike hallucinations confined to a single modality, cross-modal hallucinations reflect inadequate control over modality interactions, whereby information from one modality improperly influences the generation of content associated with another. In audio-visual settings, visual cues may induce the model to describe nonexistent sounds, while acoustic cues may lead it to describe nonexistent visual events.

Recent decoding methods mitigate multimodal hallucinations by modulating the influence of different modalities during generation. Audio-Visual Contrastive Decoding (AVCD) applies uniform perturbations to audio and visual inputs to alleviate modality dominance~\cite{AVCD}. Modality-Adaptive Decoding (MAD) estimates query-conditioned modality relevance and adaptively combines the audio-only, video-only, joint audio-visual, and text-only decoding branches~\cite{MAD}. However, as shown in Fig.~\ref{fig:Introduction}(a), neither uniform perturbation nor relevance-based branch weighting explicitly assesses the predictive compatibility of audio and visual evidence within the joint branch. Consequently, these methods do not determine whether fusion exploits useful complementarity or introduces harmful interference.

To examine this, we evaluate Qwen2.5-Omni-7B on the Visual Dom. and Audio Dom. categories of CMM~\cite{CMM}. Among samples for which both unimodal branches predict the correct answer, joint inference is incorrect in 4.5\% of cases. Conversely, when both unimodal branches are incorrect, joint inference recovers the correct answer in 8.7\% of cases. These opposing outcomes indicate that changes induced by joint fusion may reflect either destructive interference or complementary recovery.

To address this limitation, we propose \textbf{Conflict-Aware Decoding (CAD)}, a training-free framework that augments relevance-aware decoding with a two-stage conflict assessment. As illustrated in Fig.~\ref{fig:Introduction}(b), \textbf{Potential Conflict Magnitude Estimation (PCME)} introduces the unsigned score \(C\) to measure the magnitude of potential cross-modal conflict. Morever, \textbf{Conflict Actionability Assessment (CAA)} evaluates task-space answer relations using query relevance and answer decisiveness as reliability evidence. Following reliability discounting in Dempster-Shafer~\cite{Dempster-Shafer} evidence theory, CAA represents committed support and ignorance through simple-support masses. When \(C\) exceeds the intervention threshold and \(B\) provides evidence for intervention, CAD shifts weight from the joint branch to the unimodal branches and combines all four branch logits using the adjusted weights for conflict-aware generation.

Extensive experiments on CMM, AVHBench, WorldSense and VideoMME demonstrate that CAD consistently achieves higher overall accuracy than the base decoder, AVCD, and MAD. As illustrated in Fig.~\ref{fig:Introduction}(c), CAD consistently improves performance across different Omni-LLM backbones. Specifically, it increases overall accuracy on CMM and AVHBench by \(14.1\) and \(8.0\) percentage points on Qwen2.5-Omni-7B, respectively, while achieving corresponding gains of \(4.1\) and \(6.9\) percentage points on OmniVinci. Crucially, CAD achieves these improvements without model retraining.

We summarize the contributions of this paper as follows.
\begin{itemize}
    \item We distinguish modality relevance from predictive compatibility and show that audio-visual discrepancy may reflect either destructive interference or useful complementarity, motivating separate conflict-magnitude and actionability assessment.
    \item We propose CAD, a training-free decoding framework that combines PCME with a Dempster--Shafer-inspired CAA to identify actionable cross-modal conflict and selectively reallocate modality influence.
    \item We conduct extensive experiments on two cross-modal hallucination benchmarks and four audio-visual backbones. CAD achieves the highest overall accuracy for every evaluated backbone--benchmark pair over Base, AVCD, and MAD, while ablations isolate the contribution of each component.
\end{itemize}

\section{Related work}
\subsection{Omnimodal Large Language Models}
In recent years, multimodal large language models (MLLMs) have evolved from static image-text understanding toward multimodal reasoning in complex audio-visual scenarios. Conventional MLLMs~\cite{Qwen3-VL,InternVL,VideoChat,Video-LLaVA,MedGemma,video-SALMONN,LLaVA-Video} typically process different modalities in a relatively independent manner. In contrast, omnimodal large language models (Omni-LLMs)~\cite{VITA,Baichuan-Omni,Qwen3-Omni,Qwen3.5-Omni}, such as Qwen2.5-omni and OmniVinci~\cite{Qwen2.5-Omni,OmniVinci}, project text, image, video, and audio features into a unified representation space, while organizing audio and visual tokens along a shared timeline into an interleaved sequence. This unified design enables tighter cross-modal interaction and joint audio-visual reasoning. Specifically, Visual signals convey spatial layouts, object states, and motion dynamics, whereas audio provides semantic content, sound events, and temporal cues along a shared timeline~\cite{Ming-Omni,InteractiveOmni,Mini-Omni2}. Effective audio-visual reasoning therefore requires temporal modeling and complementary evidence integration; however, inconsistent signals may lead to harmful fusion conflicts.

\subsection{Mitigating Hallucinations in MLLMs}
Hallucination remains a critical challenge for MLLMs, whose responses may deviate from the evidence provided by audio, visual, or textual inputs. Existing mitigation approaches include representation debiasing~\cite{HACL,Post-HocDebias}, architectural refinement~\cite{DeCo,RoPE,Nullu,OPERA}, and training-free inference. Among them, training-free decoding methods~\cite{CD,ED} are particularly appealing because they require neither additional annotations nor parameter updates. They suppress unsupported generation through modality perturbations~\cite{VCD}, auxiliary descriptions~\cite{CODE}, instruction-level contrasts~\cite{ICD}, or layerwise distribution differences~\cite{DoLa}. However, these approaches are primarily designed for vision-language settings and therefore do not directly address hallucinations arising from interactions between audio and visual evidence in Omni-LLMs.

To address this limitation, recent studies have extended training-free decoding to audio-visual settings. AVCD~\cite{AVCD} uses attention-derived dominance scores to identify less dominant modalities and applies attentive masking to construct contrastive logits over audio, visual, and textual inputs. MAD~\cite{MAD} instead estimates query-specific modality relevance through model self-assessment and uses the resulting scores to adaptively weight modality-specific decoding branches. Despite these advances, neither method determines whether cross-modal disagreement reflects complementary evidence or a harmful fusion conflict. Differently, CAD assesses potential conflicts among audio-only, visual-only, and joint audio-visual predictions and adaptively reweights the decoding branches to reduce conflict-induced interference.

\section{Preliminaries}
\label{sec:Preliminaries}

\subsection{Modality-Adaptive Decoding}
Modality-Adaptive Decoding (MAD) is a training-free decoding method that adapts contrastive strength to the modalities required by the current query. Given audio input $X_A$, video input $X_V$, and textual query $X_Q$, MAD appends a fixed modality-query prompt $X_M$ asking whether the query relies on audio, video, or their joint use. Let $u_A$, $u_V$, and $u_{AV}$ denote the next-token logits assigned to the response tokens ``audio'', ``video'' and ``both'' respectively. The resulting category-level modality-relevance weights for audio, video, and their joint use are
\begin{equation}
(r_A,r_V,r_{AV})
=
\operatorname{softmax}
\!\left([u_A,u_V,u_{AV}]\right).
\label{eq:mad_relevance}
\end{equation}

At decoding step $t$, let $y_{<t}$ denote the generated prefix, $\mathcal V$ the model vocabulary, and $\mathcal K=\{AV,A,V,T\}$ the set of modality configurations. MAD evaluates joint audio-visual input $X^{(AV)}=(X_A,X_V)$, audio-only input $X^{(A)}=X_A$, video-only input $X^{(V)}=X_V$, and question-only input $X^{(T)}=\varnothing$. Here, $\varnothing$ denotes the absence of audio-visual context, while $X_Q$ remains in every branch. The entries of the resulting next-token logit vectors are
\begin{equation}
\begin{gathered}
z_k^t(v)
=
\operatorname{logit}_{\theta}
\!\left(
v\mid X^{(k)},X_Q,y_{<t}
\right),\\[-1mm]
k\in\mathcal K,
\qquad
v\in\mathcal V.
\end{gathered}
\label{eq:mad_branch_logits}
\end{equation}

Collecting terms in the four-branch MAD decoder yields the algebraically equivalent compact form:
\begin{equation}
\begin{aligned}
z_{\mathrm{MAD}}^t(v)
&=
\sum_{k\in\mathcal K}
w_k z_k^t(v),\\
w_{AV}
&=
2+2\gamma r_{AV},\\
w_A
&=
1-\gamma(r_{AV}-r_A),\\
w_V
&=
1-\gamma(r_{AV}-r_V),\\
w_T
&=
-\gamma(r_A+r_V).
\end{aligned}
\label{eq:mad_compact_decoding}
\end{equation}

Here, $\gamma>0$ controls the overall contrastive strength; by construction, the coefficients satisfy $\sum_{k\in\mathcal K}w_k=4$. The joint coefficient increases with joint relevance; each unimodal coefficient reflects its relevance relative to the joint configuration, while the negative question-only coefficient suppresses tokens driven primarily by the language prior. MAD then selects the next token by greedy decoding:
\begin{equation}
\hat y_t
=
\arg\max_{v\in\mathcal V}
z_{\mathrm{MAD}}^t(v).
\label{eq:mad_next_token}
\end{equation}

By weighting each modality according to its relevance to the query, MAD emphasizes task-relevant information during decoding, thereby mitigating cross-modal hallucinations.

\begin{figure}[t]
    \centering
    \includegraphics[width=0.98\linewidth]{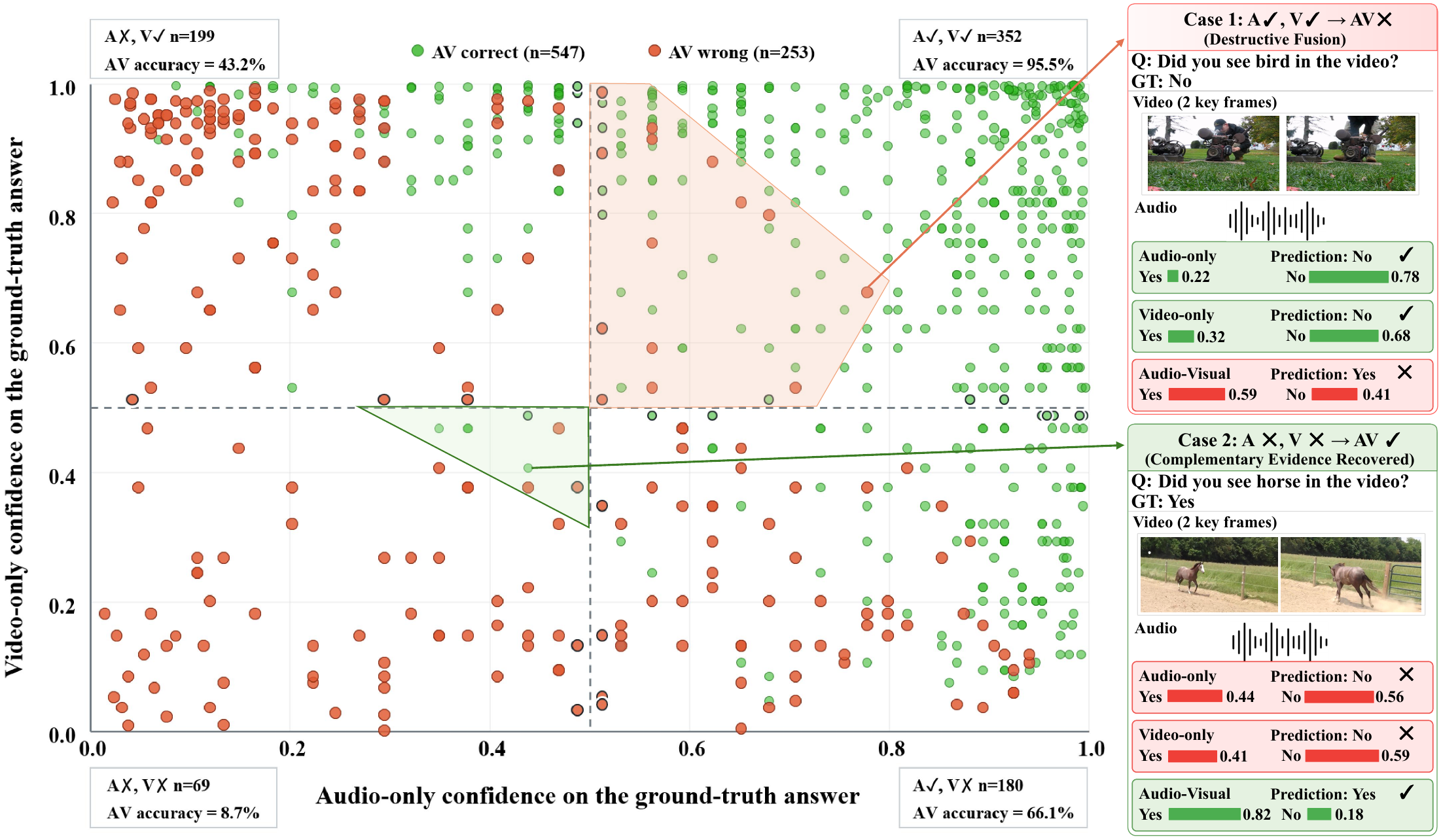}
    \caption{Fusion outcomes of Qwen2.5-Omni-7B on CMM. Axes show unimodal confidence in the ground-truth answer, and color denotes $AV$ correctness. Highlighted examples show destructive fusion and complementary recovery.}
    \label{fig:motivation}
\end{figure}

\subsection{Motivation}

To examine how multimodal fusion changes an Omni-LLM's predictions, we evaluate Qwen2.5-Omni-7B on 800 CMM questions under three input configurations: audio only (\(A\)), video only (\(V\)), and joint audio-video (\(AV\)). For this diagnostic analysis, a branch is correct when its answer-space argmax matches the ground-truth answer.

Figure~\ref{fig:motivation} reveals two opposing fusion outcomes. In the ``\(A\checkmark,V\checkmark\)'' region, AV error is \(4.5\%\). Moreover, in the ``\(A\times,V\times\)'' region, AV accuracy is \(8.7\%\). These outcomes, illustrated by the highlighted cases, show that deviation of the joint prediction from its unimodal counterparts may reflect either destructive interference or useful complementarity. 

In addition, this exposes a limitation of MAD: its branch coefficients in Eq.~\eqref{eq:mad_compact_decoding} are determined by modality relevance, indicating the query's requirements over individual modalities and their joint configuration, but do not assess whether the joint and unimodal predictions are compatible. Relevance weights alone therefore do not indicate whether an observed AV deviation reflects conflict that warrants suppression or complementarity that should be preserved. To address this, CAD separates conflict magnitude from intervention evidence: PCME measures unsigned predictive discrepancy, whereas CAA assesses whether the observed answer relation warrants intervention.

\section{Method}
\label{sec:method}

In this section, we introduce \textbf{Conflict-Aware Decoding (CAD)}. CAD extends MAD by estimating the magnitude of potential audio--visual conflict and assessing whether the detected conflict is supported by reliable evidence for intervention, thereby selectively reallocating decoding weights for actionable conflicts.

\begin{figure*}[t]
    \centering
    \includegraphics[width=\textwidth]{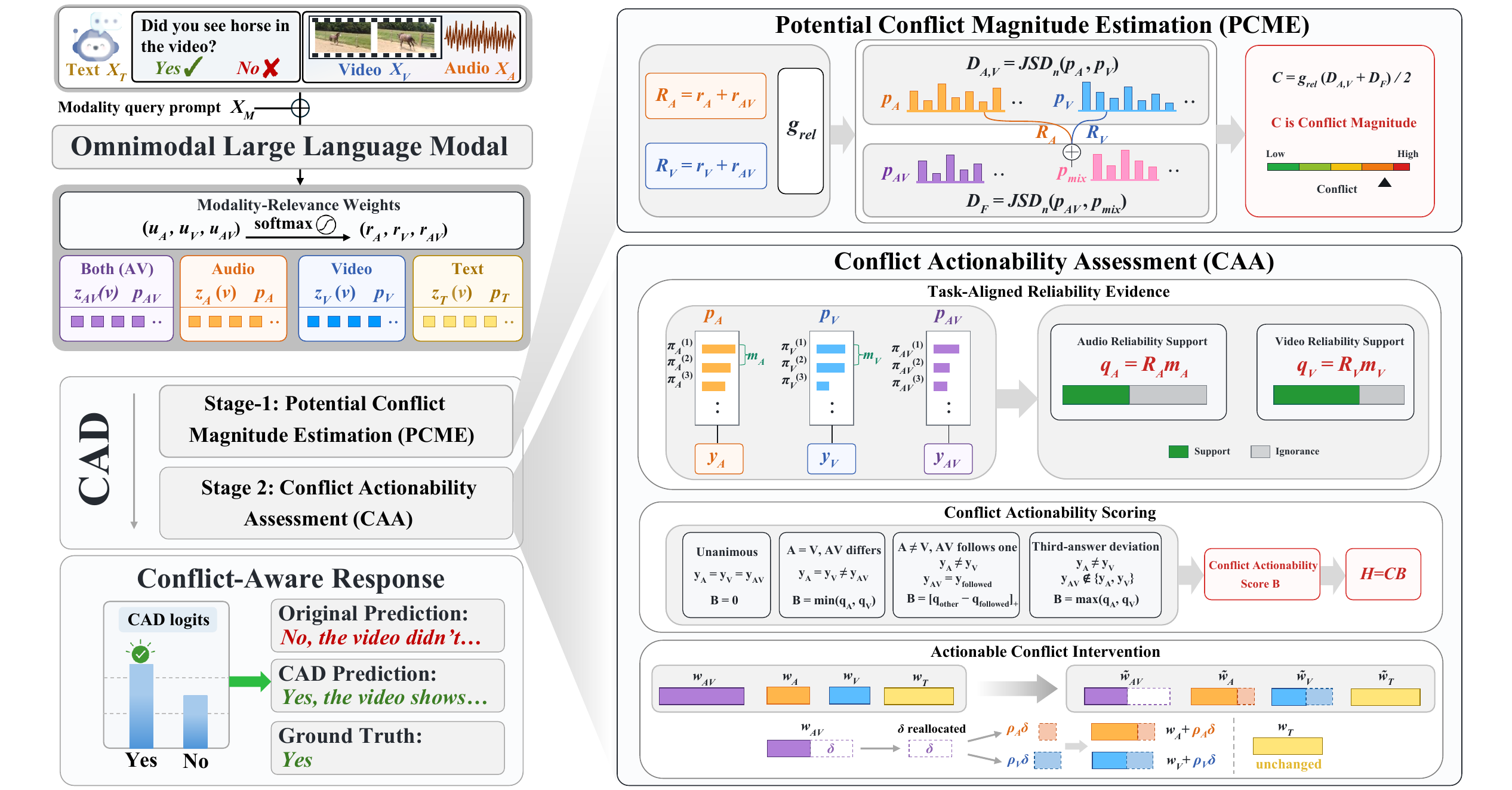}
    \caption{Overview of CAD. Given an audio-visual input and textual query, CAD uses Potential Conflict Magnitude Estimation (PCME) to measure predictive disagreement across modality branches and Conflict Actionability Assessment (CAA) to assess whether the conflict is supported by reliable unimodal evidence. The actionable-conflict signal reallocates decoding influence from the joint audio-visual branch to modality-specific branches, yielding a response better grounded in the input evidence.}
    \label{fig:overview}
\end{figure*}

\subsection{Overview of CAD}
CAD is a training-free, conflict-aware decoding method for mitigating audio-visual cross-modal hallucinations in omnimodal large language models (Omni-LLMs). As illustrated in Fig.~\ref{fig:overview}, CAD adopts a two-stage design that first estimates the magnitude of potential cross-modal conflict and then assesses whether the detected conflict is supported by reliable evidence for intervention.

In the first stage, CAD performs \textbf{Potential Conflict Magnitude Estimation (PCME)} and introduces an unsigned score $C$ that measures the magnitude of potential cross-modal conflict. $C$ captures two forms of predictive discrepancy: the disagreement between the audio-only and video-only predictions, and the deviation of the joint audio-visual prediction from a reference distribution formed by weighting the two unimodal predictions. 

In the second stage, CAD performs \textbf{Conflict Actionability Assessment (CAA)}, introducing an actionability score $B$ to quantify support for intervening on the potential cross-modal conflict identified in the first stage. Based on Dempster-Shafer evidence theory, CAD estimates modality reliability from query relevance and answer decisiveness, then combines $C$ and $B$ to identify actionable conflicts and selectively adjust decoding weights.

\subsection{Potential Conflict Magnitude Estimation}
\label{sec:conflict_magnitude}

Building on MAD's branch predictions and modality-relevance weights, Potential Conflict Magnitude Estimation (PCME) estimates the magnitude of potential cross-modal conflict by comparing the audio-only, video-only, and joint audio-visual predictions. 
First, because the question-only branch $T$ contains no audio-visual evidence, PCME excludes it from conflict estimation. For each remaining branch $k\in\{AV,A,V\}$, we convert the logit vector defined in Eq.~\eqref{eq:mad_branch_logits} into the next-token distribution
\begin{equation}
p_k^t(v)
=
\left[\operatorname{softmax}(z_k^t)\right]_v,
\qquad
v\in\mathcal V.
\label{eq:pcme_distributions}
\end{equation}

For notational simplicity, we omit the superscript $t$ hereafter. All subsequent distributions and scores refer to the same decoding step. Because the joint-use relevance weight $r_{AV}$ corresponds to a category in which both modalities are involved, each modality's effective relevance aggregates all category-level weights involving that modality:
\begin{equation}
R_A=r_A+r_{AV},
\qquad
R_V=r_V+r_{AV}.
\label{eq:effective_relevance}
\end{equation}

To focus conflict estimation on cases in which both modalities are relevant to the query, PCME introduces the relevance-balance factor
\begin{equation}
g_{\mathrm{rel}}
=
\frac{2\min(R_A,R_V)}
{R_A+R_V+\epsilon},
\label{eq:relevance_balance}
\end{equation}
where $\epsilon$ is a numerical-stability constant set to $10^{-8}$. The factor approaches one when the effective relevance scores are balanced and decreases as either modality becomes less relevant. Because $2\min(R_A,R_V)\le R_A+R_V$ and $\epsilon>0$, we have $g_{\mathrm{rel}}\in[0,1)$. PCME then constructs a relevance-weighted unimodal-branch reference:
\begin{equation}
p_{\mathrm{mix}}
=
\frac{R_Ap_A+R_Vp_V}
{\lVert R_Ap_A+R_Vp_V\rVert_1}.
\label{eq:unimodal_reference}
\end{equation}
Because audio and video may contribute unequally to the current query, this relevance-weighted combination summarizes the two unimodal predictions according to their effective relevance and serves as a reference for measuring how far the joint prediction $p_{AV}$ departs from the unimodal evidence. 

Using Jensen-Shannon divergence normalized by $\log 2$, denoted by $\operatorname{JSD}_n\in[0,1]$, PCME measures two complementary predictive discrepancies:
\begin{equation}
\begin{aligned}
D_{A,V}
&=
\operatorname{JSD}_n(p_A,p_V),\\
D_F
&=
\operatorname{JSD}_n(p_{AV},p_{\mathrm{mix}}).
\end{aligned}
\label{eq:predictive_discrepancies}
\end{equation}
Here, $D_{A,V}$ captures the disagreement between the audio-only and video-only predictions, whereas $D_F$ captures the deviation of the joint prediction from the relevance-weighted unimodal reference. 

PCME then averages the two discrepancy components without introducing an additional weighting hyperparameter and modulates their mean by the relevance-balance factor, yielding the potential-conflict magnitude
\begin{equation}
C
=
g_{\mathrm{rel}}
\frac{D_{A,V}+D_F}{2}.
\label{eq:conflict_magnitude}
\end{equation}
The resulting score $0\le C<1$ is unsigned. Because unimodal disagreement and joint-prediction deviation may reflect either beneficial cross-modal interaction or harmful interference, $C$ measures only the magnitude of potential cross-modal conflict; it neither determines whether the interaction is beneficial or harmful nor establishes that the joint prediction is erroneous. In addition, CAD considers a decoding step eligible for intervention only when $C\ge\tau$, where $\tau$ is a intervention threshold. Even when this condition is satisfied, conflict magnitude alone cannot determine whether sufficiently reliable evidence supports intervention, motivating the Conflict Actionability Assessment described next.

\subsection{Conflict Actionability Assessment}
\label{sec:conflict_actionability}

\paragraph{Task-Aligned Reliability Evidence.}
A large potential-conflict magnitude is not necessarily actionable because PCME measures Jensen--Shannon divergence over the full vocabulary. A high value can arise in two cases that do not by themselves justify intervention. First, probability may shift substantially among tokens outside the valid answer set while all branches retain the same task answer. Second, the joint branch may deviate from the unimodal distributions because it successfully integrates complementary audio and visual cues. Both cases produce distributional discrepancy without reliable evidence that the joint prediction should be suppressed. CAA therefore projects the $A$, $V$, and $AV$ predictions onto a shared task-answer space $\mathcal Y$:
\begin{equation}
\pi_k(y)=
\frac{\sum_{v\in\mathcal T(y)}p_k(v)}
{\sum_{y'\in\mathcal Y}\sum_{v\in\mathcal T(y')}p_k(v)},
\qquad
k\in\{A,V,AV\},
\label{eq:caa_answer_projection}
\end{equation}
where $\mathcal T(y)\subseteq\mathcal V$ contains the tokenizer-valid single-token surface forms of answer $y$. For multiple-choice and binary tasks, $\mathcal Y$ contains the valid options or answer labels. CAA summarizes each projected prediction by
\begin{equation}
y_k=\arg\max_{y\in\mathcal Y}\pi_k(y),
\qquad
m_k=\pi_k^{(1)}-\pi_k^{(2)},
\label{eq:caa_answer_summary}
\end{equation}
where $\pi_k^{(1)}\ge\pi_k^{(2)}$ are the two largest values of $\pi_k$. The margin $m_k$ measures answer decisiveness within the task space. Moreover, CAA discounts answer evidence by both query relevance and answer decisiveness:
\begin{equation}
q_A=R_A m_A,
\qquad
q_V=R_V m_V.
\label{eq:caa_reliability}
\end{equation}
Here, $R_k$ measures the applicability of modality $k$ to the query, while $m_k$ measures how decisively its branch selects an answer. Their product $q_k$ is an uncalibrated reliability proxy rather than a probability that $y_k$ is correct. Following the reliability-discounting principle of Dempster--Shafer theory, CAA represents each unimodal branch by the simple-support mass
\begin{equation}
M_k(\{y_k\})=q_k,
\qquad
M_k(\mathcal Y)=1-q_k,
\quad k\in\{A,V\}.
\label{eq:caa_simple_support}
\end{equation}
The committed mass expresses reliability-discounted support for $y_k$, and the remaining mass expresses ignorance. Unlike independent physical sensors, the audio and video branches share the model backbone, textual query, and language prior. Directly applying Dempster's rule may therefore double-count shared evidence. We adopt a conservative adaptation: the actionability rule requires support from both agreeing branches through $\min(q_A,q_V)$ and, when the unimodal branches disagree and the joint branch follows one of them, retains only the bounded positive reliability advantage of the opposing branch. This preserves the Dempster-Shafer principle of separating committed support from ignorance while adapting evidence aggregation to dependent model branches. These quantities provide operational evidence for intervention, with $C$ measuring distributional conflict via JSD rather than Dempster's conflict coefficient.

\paragraph{Conflict Actionability Scoring.}
CAA maps the answer relations and reliability proxies to
\begin{equation}
B=
\begin{cases}
0,
& y_A=y_V=y_{AV},\\
\min(q_A,q_V),
& y_A=y_V\ne y_{AV},\\
[q_{\bar{k}}-q_k]_+,
& y_A\ne y_V,\ y_{AV}=y_k,\\
\max(q_A,q_V),
& y_{AV}\notin\{y_A,y_V\},
\end{cases}
\label{eq:caa_answer_gate}
\end{equation}
where $[x]_+=\max(x,0)$. Here, $k\in{A,V}$ denotes the unimodal branch followed by $AV$, and $\bar{k}$ the other branch. When all branches agree, no evidence supports intervention, so $B=0$ even if their full-vocabulary distributions differ. When the unimodal branches agree but $AV$ differs, intervention is supported only by their shared reliability, yielding the conservative minimum. When the unimodal answers differ and $AV$ follows branch $k$, support is positive only if the competing branch provides stronger reliable evidence. When $AV$ produces a third answer, CAA uses the strongest unimodal opposition, $\max(q_A,q_V)$, as a conservative heuristic. This case is not a strict Dempster derivation and does not imply that either unimodal answer is correct. Across all cases, $B$ measures how strongly the observed answer relation warrants intervention; it neither labels the interaction as harmful nor identifies the ground-truth answer. CAA then forms the actionable-conflict signal
\begin{equation}
H=CB.
\label{eq:caa_actionable_conflict}
\end{equation}
Because $0\le B\le1$ and $0\le C<1$, we have $0\le H\le C<1$. $H$ is a CAD-specific operational score requiring both conflict magnitude and evidential support for intervention. A high value therefore requires both substantial distributional conflict and reliable support for intervention.

\paragraph{Actionable Conflict Intervention.}
Using the base MAD coefficients defined in
Eq.~\eqref{eq:mad_compact_decoding}, let
$s=[w_A]_+ + [w_V]_+$. CAD allocates the transferred influence in
proportion to the positive unimodal coefficients:
\begin{equation}
(\rho_A,\rho_V)=
\begin{cases}
([w_A]_+/s,[w_V]_+/s), & s>0,\\
(1/2,1/2), & s=0.
\end{cases}
\label{eq:caa_allocation_ratio}
\end{equation}
For intervention strength $\lambda>0$, the transfer amount is
\begin{equation}
\delta=
\begin{cases}
\lambda H[w_{AV}]_+, & C\ge\tau,\\
0, & C<\tau.
\end{cases}
\label{eq:caa_transfer}
\end{equation}
The adjusted coefficients are
\begin{equation}
\begin{aligned}
\widetilde w_{AV}&=w_{AV}-\delta,\\
\widetilde w_A&=w_A+\rho_A\delta,\\
\widetilde w_V&=w_V+\rho_V\delta,\\
\widetilde w_T&=w_T.
\end{aligned}
\label{eq:caa_adjusted_weights}
\end{equation}
Since $\rho_A+\rho_V=1$, CAD shifts decoding influence from the joint branch to the unimodal branches while preserving the coefficient sum and leaving $w_T$ unchanged. If $C<\tau$ or $B=0$, then $\delta=0$, reducing CAD to the base MAD decoder. When $\lambda H>1$, $\widetilde w_{AV}<0$, yielding contrastive subtraction of the joint branch. At decoding step $t$, CAD selects
\begin{equation}
\hat y_t=
\arg\max_{v\in\mathcal V}
\sum_{k\in\{AV,A,V,T\}}
\widetilde w_k z_k^t(v).
\label{eq:caa_final_decoding}
\end{equation}

CAD computes $C$, $B$, and the adjusted coefficients from the initial next-token distributions and keeps the coefficients fixed throughout generation.

\section{Experiments}
\subsection{Experimental Setting}
\paragraph{Benchmarks.}
We evaluate our method on four benchmarks. AVHBench~\cite{AVHBench} and CMM~\cite{CMM} specifically assess cross-modal hallucinations in audio-visual large language models. AVHBench focuses on video-driven audio hallucinations and audio-driven video hallucinations, while CMM evaluates modality dominance and cross-modal interference across diverse audio-visual scenarios. We further use WorldSense~\cite{WorldSense} and VideoMME~\cite{VideoMME} to assess general audio-visual question answering. Together, these benchmarks evaluate both hallucination mitigation and general audio-visual reasoning ability.

\paragraph{Comparison Methods.}
We compare CAD with two training-free decoding methods: AVCD~\cite{AVCD} and MAD~\cite{MAD}. AVCD extends contrastive decoding to audio-visual models but applies uniform modality perturbations without query-specific adaptation. MAD dynamically adjusts multimodal predictions according to the relevance of different modalities, thereby mitigating cross-modal hallucinations.

\begin{table*}[t]
    \centering

    {
    \small

    \setlength{\tabcolsep}{1mm}

    \renewcommand{\arraystretch}{0.88}

    \begin{tabularx}{\textwidth}{
        @{}
        >{\raggedright\arraybackslash}X
        C{1.70cm}
        C{1.70cm}
        C{1.70cm}
        C{2.05cm}
        C{2.05cm}
        C{2.05cm}
        @{}
    }
        \toprule

        \multirow[c]{2}{*}{\textbf{Model}}
        & \multicolumn{3}{c}{\textbf{CMM}}
        & \multicolumn{3}{c}{\textbf{AVHBench}} \\

        \cmidrule(lr){2-4}
        \cmidrule(lr){5-7}

        &
        \makecell[c]{\textbf{Visual Dom.}}
        &
        \makecell[c]{\textbf{Audio Dom.}}
        &
        \makecell[c]{\textbf{Overall Acc.}}
        &
        \makecell[c]{\textbf{Video-Driven}\\\textbf{Audio Hall.}}
        &
        \makecell[c]{\textbf{Audio-Driven}\\\textbf{Video Hall.}}
        &
        \makecell[c]{\textbf{Overall Acc.}}
        \\

        \midrule

        VideoLLaMA2-AV
        & 72.3
        & 79.5
        & 75.9
        & 75.6
        & 78.8
        & 76.6 \\

        VideoLLaMA2-AV + AVCD
        & 72.5
        & 81.8
        & 77.1
        & 72.1
        & 76.8
        & 73.6 \\

        VideoLLaMA2-AV + MAD
        & 81.8
        & 82.3
        & 82.0
        & 78.5
        & 80.4
        & 79.7 \\

        \rowcolor{gray!15}
        VideoLLaMA2-AV + CAD
        & \textbf{84.5}
        & \textbf{82.3}
        & \textbf{83.4}
        & \textbf{81.3}
        & \textbf{80.6}
        & \textbf{81.1} \\

        \midrule

        OmniVinci
        & 82.3
        & 85.5
        & 83.9
        & 74.5
        & 71.2
        & 73.4 \\

        OmniVinci + AVCD
        & 81.3
        & 87.0
        & 84.1
        & 70.9
        & 67.0
        & 69.6 \\

        OmniVinci + MAD
        & 83.8
        & 87.5
        & 85.6
        & 78.0
        & 73.9
        & 76.7 \\

        \rowcolor{gray!15}
        OmniVinci + CAD
        & \textbf{87.0}
        & \textbf{89.0}
        & \textbf{88.0}
        & \textbf{82.6}
        & \textbf{75.7}
        & \textbf{80.3} \\
        
        \midrule

        Qwen2.5-Omni-3B
        & 69.3
        & 82.5
        & 75.9
        & 74.3
        & 80.2
        & 76.2 \\

        Qwen2.5-Omni-3B + AVCD
        & 75.3
        & 78.0
        & 76.6
        & 73.5
        & 78.7
        & 75.2 \\

        Qwen2.5-Omni-3B + MAD
        & 76.0
        & 87.3
        & 81.6
        & 80.2
        & 82.1
        & 80.8 \\

        \rowcolor{gray!15}
        Qwen2.5-Omni-3B + CAD
        & \textbf{77.3}
        & \textbf{88.5}
        & \textbf{82.9}
        & \textbf{82.6}
        & \textbf{82.2}
        & \textbf{82.5} \\

        \midrule

        Qwen2.5-Omni-7B
        & 69.5
        & 72.3
        & 70.9
        & 75.7
        & 76.9
        & 76.1 \\

        Qwen2.5-Omni-7B + AVCD
        & 66.3
        & 70.8
        & 68.5
        & 75.6
        & 77.0
        & 76.1 \\

        Qwen2.5-Omni-7B + MAD
        & 77.3
        & 84.0
        & 80.6
        & 79.4
        & 84.7
        & 81.2 \\

        \rowcolor{gray!15}
        Qwen2.5-Omni-7B + CAD
        & \textbf{83.3}
        & \textbf{86.8}
        & \textbf{85.0}
        & \textbf{83.2}
        & \textbf{85.9}
        & \textbf{84.1} \\

        \bottomrule
    \end{tabularx}
    }
    \caption{
    Main results on cross-modal hallucination benchmarks.
    We compare the original decoding strategy (Base),
    AVCD, MAD, and our proposed CAD.
    All results are reported in accuracy (\%).
    }
    \label{tab:main_results}
\end{table*}

\paragraph{Implementation Details.}
We evaluate CAD on four representative audio-visual large language models: VideoLLaMA2-AV~\cite{VideoLLaMA2-AV}, OmniVinci~\cite{OmniVinci}, Qwen2.5-Omni-3B, and Qwen2.5-Omni-7B~\cite{Qwen2.5-Omni}. All experiments are conducted on NVIDIA L20 GPUs with 48 GB of memory. Following prior work, we sample videos from CMM, AVHBench, and VideoMME at 2 fps, with the maximum number of sampled frames capped at 768. For WorldSense, we use a sampling rate of 1 fps and cap the number of frames at 384. Regarding hyperparameters, the intervention threshold $\tau$ is set to 0.05. We use the same computational setup across all experiments to ensure reproducibility.

\subsection{Main Results}
Table~\ref{tab:main_results} Table~\ref{tab:main_results} compares CAD with competing decoding methods on CMM and AVHBench. CAD consistently improves performance across backbones and hallucination categories, demonstrating its effectiveness in mitigating cross-modal interference.

\paragraph{Results on CMM.}
CMM evaluates hallucinations arising from overreliance on unimodal priors, including visual dominance, where visual cues overshadow relevant auditory and linguistic evidence, and audio dominance, where auditory cues suppress relevant visual or linguistic information.

CAD consistently improves performance across all evaluated backbones. On Qwen2.5-Omni-7B, it increases visual- and audio-dominance accuracy by 13.8 and 14.5 percentage points over the base model, respectively, raising the overall accuracy by 14.1 points to 85.0\%. Similar improvements on VideoLLaMA2-AV, OmniVinci, and Qwen2.5-Omni-3B further demonstrate the effectiveness of CAD in mitigating hallucinations caused by unimodal bias.

\paragraph{Results on AVHBench.}
CAD also yields strong gains on AVHBench. On Qwen2.5-Omni-7B, it improves video-driven audio and audio-driven video hallucination accuracy by 7.5 and 9.0 percentage points, respectively, raising the overall accuracy by 8.0 points to 84.1\%. Qwen2.5-Omni-3B and OmniVinci show gains of 8.3 and 8.1 points in the video-driven audio setting, respectively, while VideoLLaMA2-AV improves in both categories. These results confirm that CAD effectively mitigates bidirectional cross-modal hallucinations across different backbones.

Compared with AVCD’s uniform modality perturbation and MAD’s relevance-based adaptation, CAD further assesses whether joint audio-visual predictions are supported by modality-specific evidence. Its conflict-aware reallocation suppresses unreliable fusion while preserving valid cross-modal interactions, highlighting the importance of explicit conflict modeling for reliable omnimodal reasoning

\begin{table}[t]
    \centering

    {
    \small

    \setlength{\tabcolsep}{1mm}

    \begin{tabularx}{\linewidth}{@{}cYYY@{}}
        \toprule

        \multirow[c]{2}{*}[-1.3ex]{%
            \makecell[c]{\textbf{Fusion}\\\textbf{Method}}%
        }
        & \multicolumn{3}{c}{\textbf{AVHBench}} \\

        \cmidrule(lr){2-4}

        & \textbf{\makecell[c]{Video-Driven\\Audio Hall.}}
        & \textbf{\makecell[c]{Audio-Driven\\Video Hall.}}
        & \textbf{Overall Acc.} \\

        \midrule

        Baseline
        & 75.7
        & 76.9
        & 76.1 \\

        Uniform
        & 75.9
        & 82.2
        & 78.0 \\

        Argmax
        & 78.6
        & 81.8
        & 79.6 \\

        \rowcolor{gray!15}
        CAD
        & \textbf{83.2}
        & \textbf{85.9}
        & \textbf{84.1} \\

        \bottomrule
    \end{tabularx}
    }
    \caption{Comparison of different modality fusion strategies on AVHBench using Qwen2.5-Omni-7B. CAD achieves the highest overall accuracy among all strategies.}
    \label{tab:fusion_ablation}
\end{table}

\paragraph{Comparison of Weighting Strategies.}
To examine the contributions of modality-relevance estimation and conflict-aware intervention, we compare CAD with two simplified weighting strategies. Uniform removes query-specific preferences by setting $r_A = r_V = r_{AV} = 1/3$, while Argmax converts the soft relevance distribution into a one-hot assignment. In both variants, the coefficients $w_A$, $w_V$, $w_{AV}$, and $w_T$ are directly used to compute $\hat{y}_t$ without conflict estimation or reallocation. As shown in Table~\ref{tab:fusion_ablation}, both outperform the base model but remain inferior to CAD. Uniform reaches 78.0\% overall accuracy, indicating that fixed weighting cannot capture query-specific evidence requirements, whereas Argmax reaches 79.6\% but discards complementary information from non-selected modalities. By combining soft relevance-aware weighting with conflict-aware reallocation, CAD achieves 84.1\% overall accuracy, confirming the complementary value of both components.

\begin{table}[t]
    \centering

    \resizebox{\columnwidth}{!}{%
        \begin{tabular}{
            c
            *{3}{C{1.35cm}}
            c
            c
            c
        }
            \toprule

            \multirow[c]{2}{*}{%
                \raisebox{-1.0ex}[0pt][0pt]{\textbf{Decoding}}%
            }
            & \multicolumn{3}{c}{\textbf{Modality-Relevance Weights}}
            & \multicolumn{2}{c}{\textbf{AVHBench}}
            & \multirow[c]{2}{*}{%
                \raisebox{-1.0ex}[0pt][0pt]{\textbf{Acc$\uparrow$}}%
            } \\

            \cmidrule(lr){2-4}
            \cmidrule(lr){5-6}

            & $r_A$
            & $r_V$
            & $r_{AV}$
            & \makecell[c]{\textbf{Video-Driven}\\\textbf{Audio Hall.}}
            & \makecell[c]{\textbf{Audio-Driven}\\\textbf{Video Hall.}}
            & \\

            \midrule

            CAD
            &
            & \checkmark
            & \checkmark
            & 80.6
            & 85.8
            & 82.3 \\

            CAD
            & \checkmark
            &
            & \checkmark
            & 81.4
            & 80.0
            & 80.9 \\

            CAD
            & \checkmark
            & \checkmark
            &
            & 82.3
            & 84.9
            & 83.2 \\

            CAD
            & \checkmark
            & \checkmark
            & \checkmark
            & \textbf{83.2}
            & \textbf{85.9}
            & \textbf{84.1} \\

            \bottomrule
        \end{tabular}%
    }
    \caption{
        Ablation of modality-relevance weights in CAD on AVHBench using Qwen2.5-Omni-7B. We set each of $r_A$, $r_V$, and $r_{AV}$ to zero before coefficient construction. Using all three weights yields the highest overall accuracy.
    }
    \label{tab:weight_ablation}
\end{table}

\paragraph{Contribution of Modality-Relevance Weights.}
To assess the role of modality-relevance estimation in CAD, we individually set $r_A$, $r_V$, or $r_{AV}$ to zero before constructing the decoding coefficients, while keeping the remaining weights and the subsequent conflict-aware procedure unchanged. As shown in Table~\ref{tab:weight_ablation}, removing any relevance weight degrades performance. Excluding $r_A$ mainly reduces video-driven audio hallucination accuracy, indicating that audio relevance helps prevent visual evidence from dominating audio-related predictions. Removing $r_V$ causes the largest degradation, lowering audio-driven video hallucination accuracy from 85.9\% to 80.0\% and overall accuracy from 84.1\% to 80.9\%, which highlights the importance of visual relevance in resisting misleading auditory cues. Removing $r_{AV}$ produces smaller but consistent drops in both categories, showing that joint relevance contributes to coordinating complementary audio-visual evidence. The full model achieves the best performance, confirming that the three relevance weights provide complementary guidance for conflict-aware decoding.

\begin{table}[t]
    \centering

    \resizebox{\columnwidth}{!}{%
        \begin{tabular}{
            c
            >{\centering\arraybackslash}m{1.3cm}
            >{\centering\arraybackslash}m{1.3cm}
            c
            c
            c
        }
            \toprule

            \multirow[c]{2}{*}{%
                \raisebox{-1.0ex}[0pt][0pt]{\textbf{Decoding}}%
            }
            & \multicolumn{2}{c}{\textbf{Conflict Assessment}}
            & \multicolumn{2}{c}{\textbf{AVHBench}}
            & \multirow[c]{2}{*}{%
                \raisebox{-1.0ex}[0pt][0pt]{\textbf{Acc$\uparrow$}}%
            } \\

            \cmidrule(lr){2-3}
            \cmidrule(lr){4-5}

            & \multicolumn{1}{c}{$C$}
            & \multicolumn{1}{c}{$B$}
            & \makecell[c]{%
                \textbf{Video-Driven}\\
                \textbf{Audio Hall.}%
            }
            & \makecell[c]{%
                \textbf{Audio-Driven}\\
                \textbf{Video Hall.}%
            }
            & \\

            \midrule

            CAD
            &
            & 
            & 78.7
            & 84.9
            & 80.7 \\

            CAD
            & \checkmark
            &
            & 81.5
            & 85.0
            & 82.7 \\

            CAD
            & 
            & \checkmark
            & 82.3
            & 85.5
            & 83.3 \\
            
            CAD
            & \checkmark
            & \checkmark
            & \textbf{83.2}
            & \textbf{85.9}
            & \textbf{84.1} \\

            \bottomrule
        \end{tabular}%
    }
    \caption{
        Ablation of conflict-aware components in CAD on AVHBench using Qwen2.5-Omni-7B. $C$ denotes potential cross-modal conflict magnitude, and $B$ the reliability-weighted conflict-direction gate.
    }
    \label{tab:components_ablation}
\end{table}

\paragraph{Contribution of Conflict-Aware Components.}
To isolate the contributions of conflict magnitude and directional evidence, we compare four CAD variants. The first directly decodes with the relevance-conditioned coefficients $w_A$, $w_V$, $w_{AV}$, and $w_T$ without conflict-aware intervention. The second removes the direction gate $B$ and uses $H = C$, while the third sets $C = 1$ and $\tau = 0$, such that $H = B$. The full model adopts $H = CB$. As shown in Table~\ref{tab:components_ablation}, introducing $C$ improves overall accuracy from 80.7\% to 82.7\%, while using $B$ alone achieves 83.3\%. Adding both $C$ and $B$ further increases overall accuracy to 84.1\% and improves both hallucination categories. This suggests that $C$ and $B$ provide complementary signals for conflict-aware intervention. Overall, $C$ identifies potentially problematic fusion, whereas $B$ determines whether the detected discrepancy warrants correction, resulting in more accurate conflict-aware adjustment.

\begin{figure}[t]
    \centering
    \includegraphics[width=1\linewidth]{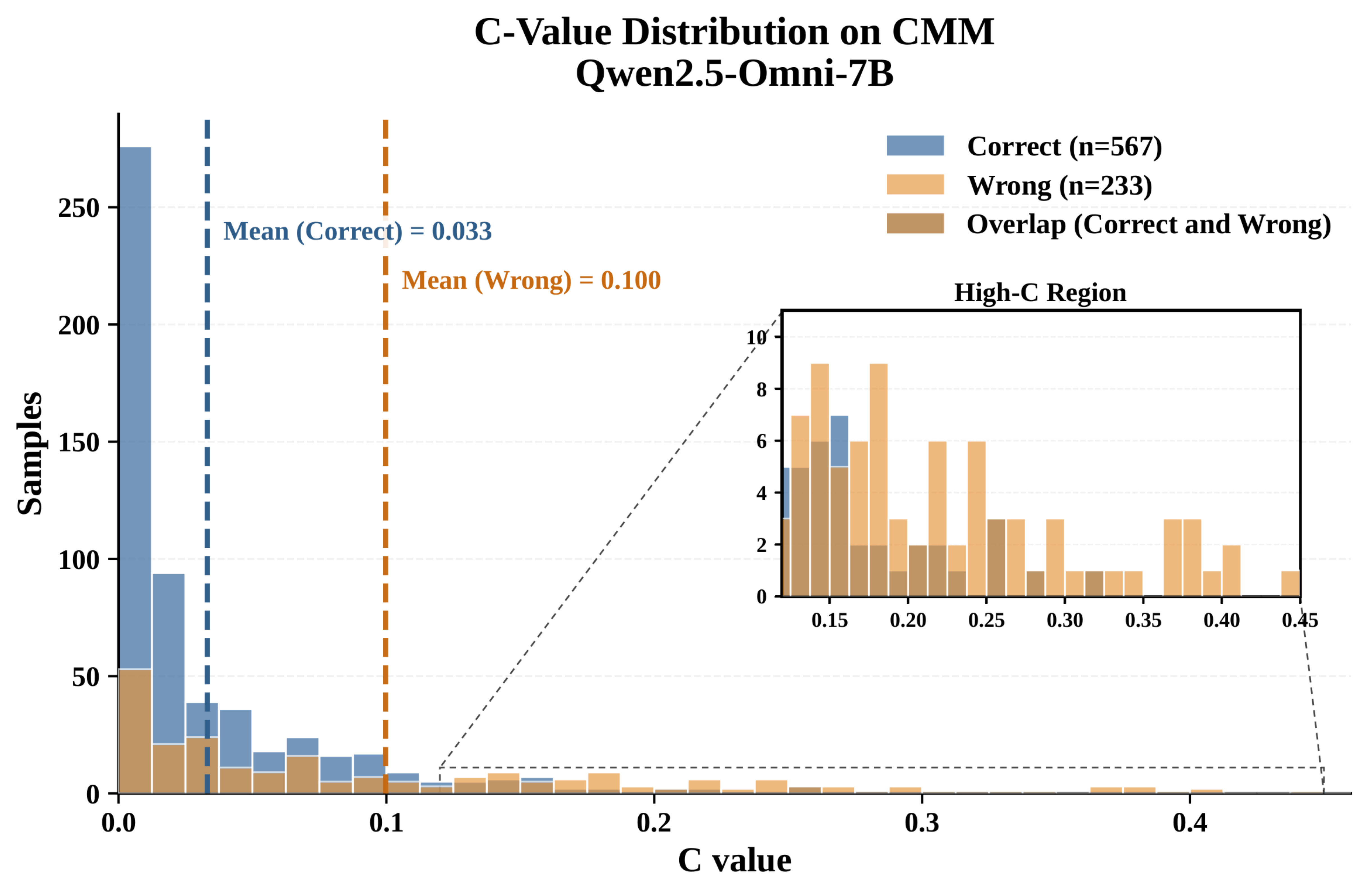}
    \caption{Distribution of potential conflict magnitude $C$ for correct and wrong Qwen2.5-Omni-7B predictions on CMM.}
    \label{fig:C_value}
\end{figure}

\paragraph{Effectiveness of Potential Conflict Magnitude.}
To assess whether $C$ captures unreliable cross-modal fusion, we compare its distributions for correct and wrong predictions of Qwen2.5-Omni-7B on CMM. As shown in Fig.~\ref{fig:C_value}, the mean $C$ for wrong predictions reaches 0.100, more than three times the 0.033 observed for correct predictions. Wrong predictions are also more concentrated in the high-$C$ region, indicating that stronger disagreement between modality-specific and joint predictions is closely associated with cross-modal reasoning errors. Overall, the clear distributional difference demonstrates that $C$ effectively characterizes the severity of potential cross-modal conflict.

\begin{table}[htbp]
    \centering

    {
    \small

    \setlength{\tabcolsep}{1mm}

    \begin{tabularx}{\columnwidth}{
        @{}
        >{\raggedright\arraybackslash}X
        c
        c
        @{}
    }
        \toprule

        \textbf{Model}
        & \textbf{Worldsense}
        & \textbf{VideoMME} \\

        \midrule

        VideoLLaMA2-AV
        & 25.9
        & 47.9 \\

        VideoLLaMA2-AV + MAD
        & 25.0
        & 48.5 \\

        \rowcolor{gray!15}
        VideoLLaMA2-AV + CAD
        & \textbf{28.8}
        & \textbf{49.4} \\

        \bottomrule
    \end{tabularx}
    }
    \caption{Comparison of Omni-LLMs on general AVQA benchmarks. All results are reported in accuracy (\%).}
    \label{tab:avqa_general}
\end{table}

\paragraph{Performance on General Audio-Visual QA Tasks.}
To evaluate whether CAD also benefits general audio-visual question answering beyond hallucination-focused benchmarks, we further assess it on WorldSense and VideoMME (w/o Sub.). As shown in Table~\ref{tab:avqa_general}, CAD achieves the best performance on both benchmarks, consistently outperforming the base model and MAD. We attribute these gains to its ability to suppress unreliable cross-modal interactions while preserving complementary audio-visual evidence. These results suggest that conflict-aware decoding not only mitigates explicit cross-modal hallucinations, but also generalizes effectively to audio-visual question answering.

\section{Conclusion}
In this work, we propose Conflict-Aware Decoding (CAD), a training-free framework for mitigating cross-modal hallucinations in Omni-LLMs. CAD assesses the magnitude and actionability of potential conflicts among audio, visual, and joint audio-visual predictions, then selectively reallocates decoding weights to suppress unreliable fusion. Experiments on CMM and AVHBench across four representative audio-visual models demonstrate consistent improvements over strong training-free baselines for both video-driven audio and audio-driven video hallucinations. Although CAD effectively reduces cross-modal hallucinations, its current implementation relies on manually tuned hyperparameters, which may limit its adaptability across models and datasets. Developing adaptive parameter-selection strategies therefore represents a promising direction for future work.

\bibliography{CAD_arxiv}

\end{document}